\documentclass[useAMS,usenatbib,onecolumn]{mn2e}
\usepackage{epsfig,amssymb}

\def\aa{{\it Astron. Astrophys.} \,}

\def\apj{{\it ApJ \,}}
\def\apjs{{\it Ap. J. Supp.} \,}

\def\jca{{\it J. Cosmo. Astro. Phys.} \,}

\def\mn{{\it MNRAS} \,}

\title[Cluster formation times and Einstein Radii]{The probability distribution of cluster formation times\\
and implied Einstein Radii}
\author[S. Sadeh, Y. Rephaeli]{Sharon Sadeh$^{1}$\thanks{E-mail:
shrs@post.tau.ac.il}, Yoel Rephaeli$^{1,2}$\\
$^{1}$School of Physics and Astronomy, Tel Aviv University, Tel Aviv, 69978,
Israel\\
$^{2}$Center for Astrophysics and Space Sciences, University of California,
San Diego, La Jolla, CA 92093-0424\\}

\begin{document}
\pagerange{\pageref{firstpage}--\pageref{lastpage}} \pubyear{2008}

\maketitle
\label{firstpage}

\begin{abstract}
We provide a quantitative assessment of the probability distribution 
function of the concentration parameter of galaxy clusters. We do so by 
using the probability distribution function of halo formation times, 
calculated by means of the excursion set formalism, and a formation 
redshift-concentration scaling derived from results of N-body simulations. 
Our results suggest that the observed high concentrations of several 
clusters are quite unlikely in the standard $\Lambda$CDM cosmological 
model, but that due to various inherent uncertainties, the statistical 
range of the predicted distribution may be significantly wider than 
commonly acknowledged. In addition, the probability distribution function 
of the Einstein radius of A1689 is evaluated, confirming that the observed 
value of $\sim 45" \pm 5"$ is very improbable in the currently favoured 
cosmological model. If, however, a variance of $\sim 20\%$ in the 
theoretically predicted value of the virial radius is assumed, then 
the discrepancy is much weaker. The measurement of similarly large 
Einstein radii in several other clusters would pose a difficulty to 
the standard model. If so, earlier formation of the large scale structure 
would be required, in accord with predictions of some quintessence models.  
We have indeed verified that in a viable early dark energy model large 
Einstein radii are predicted in as many as a few tens of high-mass clusters.

\end{abstract}

\section{Introduction}

The formation of galaxies and their systems (`haloes') is known to be more 
intricate than its simplified rendering in the context of spherical 
collapse models. The process is characterised by gradual growth of the 
system mass and evolution of its morphological properties through 
multiple merging events, evidence for which comes from observations and 
hydrodynamical simulations. In clusters, mergers affect also the 
evolution of intracluster (IC) gas density and temperature and their 
spatial profiles. On the theoretical side, studies of halo mergers and 
related issues began with the works of Bond et al. (1991) and Lacey \& 
Cole (1993), who developed the theory of \emph{excursion sets} in the 
context of structure formation. This approach was originally devised by 
Bond et al. in order to address the ``cloud-in-cloud'' problem, who 
showed that the Press \& Schechter (1974) mass function, including the 
``fudge factor'' of 2, could be derived under certain assumptions. They 
also used their formalism to derive expressions for merger probabilities. 
Lacey \& Cole (hereafter LC) used the excursion set formalism (ESF) to 
extract such related quantities as halo merger rates, halo survival 
times, and halo formation times.

The NFW concentration parameter (Navarro, Frenk, \& White 1995), which 
characterises dark matter (DM) distribution in a halo is known from N-body 
simulations to be correlated with its formation time (e.g. Jing 2000; 
Bullock et al. 2001; Zhao et al. 2003) due to the fact that haloes which 
form earlier are likely to have more condensed cores, reflecting the 
higher background density of the universe. We use this inferred 
correlation, and the probability distribution function (PDF) of halo 
formation times, to construct PDFs of halo concentrations. This provides 
a convenient framework for a comparison of theoretically predicted values 
of the concentration parameter to observational results. Our calculations 
of the formation time and concentration parameter PDFs are carried out in 
a standard $\Lambda$CDM cosmology, adopting the set of cosmological 
parameters,($\Omega_m, \Omega_{\Lambda},h,n,\sigma_8)=(0.3\pm 0.021,
0.7\pm 0.021,0.687\pm 0.018, 0.953\pm 0.016,0.827^{+0.026}_{-0.025})$, 
extracted from the 3-year WMAP+WL data (Spergel et al. 2007). These 
parameters are selected for the sake of consistency with results 
obtained in other works, including calculations of the PDF of Einstein 
radii. Specifically, we show that the observed high Einstein radius in the lensing cluster A1689, whose value is roughly in the range $40" - 50"$ (e.g., Broadhurst \& Barkanna 2008), is very improbable in the standard $\Lambda$CDM cosmology, but much more probable in a cosmological model characterised by an early component of dark energy, provided that its virial radius is larger (within a plausible range of variance) than what is predicted in the simple spherical collapse scenario.
 
This paper is arranged as follows: In \S 2 we briefly detail the derivation of the PDFs of halo formation time and concentration 
parameter; results are provided in \S 3. The case 
of the lensing cluster A1689 is explored in \S 4, 
followed by a discussion, \S 5.

\section{Method}
\subsection{PDF of formation times}
\label{sec:formt}

In the LC formalism the cumulative distribution function (CDF) of halo 
formation times, which describes the probability that a halo of mass 
$M_2$ had a parent with mass in the range [$f M_2< M_1< M_2$] at redshift 
$z_1$, with $f$ denoting the fraction of mass assembled through mergers by 
redshift $z_1$ is, 
\begin{equation}
P(M_1>f M_2,z_1|M_2,z_2)=
\int^{S_h}_{S_2}\frac{M(S_2)}{M(S_1)}
f_{S_1}(S_1,\delta_{c_1}|S_2,\delta_{c_2})\,dS_1.
\label{eq:cdf}
\end{equation}
Here $S\equiv\sigma^2$ and $\delta_c$ denote the mass variance and 
critical density for spherical collapse, respectively, with $\delta_c$ 
extrapolated to $z=0$, and $f_{S_1}(S_1,\delta_{c_1}|S_2,\delta_{c_2})$ 
is the conditional probability that a halo with mass $M_2$ at time $t_2$ 
had a progenitor of mass $M_1$ at time $t_1$. Put in the language of the 
ESF, this is the probability that a random walk trajectory which reached 
the $\delta_{c_2}$ barrier at mass scale $M_2$ had traversed for the first 
time the $\delta_{c_1}$ barrier at mass scale $M_1$. 
This probability is the key result from the ESF which is used to determine 
the halo mass function and quantities related to its hierarchical 
evolution, such as the PDF of formation times. For primordial Gaussian 
density fields,
\begin{equation}
f_{S_1}(S_1,\delta_{c_1}|S_2,\delta_{c_2})dS_1=
\frac{(\delta_{c_1}-\delta_{c_2})}{(2\pi)^{1/2}(S_1-S_2)^{3/2}}
\exp\left[-\frac{(\delta_{c_1}-\delta_{c_2})^2}{2(S_1-S_2)}\right]dS_1.
\end{equation}

This analytical result derived from the ESF is an outcome of the fact that 
smoothed Gaussian fields still obey a Gaussian distribution, and inherently 
includes the "fudge factor" of 2 which Press \& Schechter have "artificially" introduced into their mass function so as to ensure that all the mass is included in haloes. The PDF of halo formation times can be now obtained by differentiating equation~(\ref{eq:cdf}) with respect to the redshift:
\begin{equation}
p_z(z)=-\frac{\partial P(M_1>fM_2,z1|M_2,z_2)}{\partial z}.
\label{eq:pz}
\end{equation}
The definition of halo formation time is somewhat arbitrary; haloes are 
commonly considered to have formed when a fraction $f$ of their total mass 
was assembled, usually with either $f=0.5$ or $f=0.75$. It is important to 
make the distinction between the observation and formation times: The former 
corresponds to the redshift at which the halo is observed to be; the 
formation redshift is distributed in the range $(z_{obs},\infty)$, and 
its PDF must satisfy
\begin{equation}
\int_{z_{obs}}^{\infty}p_z(z)\,dz=1.
\end{equation}

\subsection{PDF of halo concentrations}

The concentration parameter is defined as $c_{v}=R_{v}/R_s$, where $R_{v}$ 
is the virial radius of the halo and $R_s$ is a characteristic inner radius, 
which roughly marks the transition radius from a $\sim 1/r$ to $\sim 1/r^3$ 
behaviour of the NFW density profile (Navarro, Frenk, \& White 1995). 
Results of N-body simulations suggest that haloes formed at higher redshifts 
tend to be associated with higher concentration parameters. This has been 
explained as an outcome of the formation of higher density cores when the 
background density was higher. This association directly implies also a 
correlation between the concentration parameter and formation time, 
evidence for which has been seen in N-body simulations of DM haloes 
(e.g. Zhao et al. 2003, Wechsler et al. 2002). 

An analytic expression describing the correlation between the concentration 
parameter and formation time can be used in order to derive a PDF of halo 
concentrations, yielding the probability of finding a given concentration 
parameter for a given mass, or the CPDF of finding a concentration higher 
than a given value. Such an expression has been deduced (Wechsler et al. 
2002) from a statistical sample of haloes identified in N-body simulations, 
\begin{equation}
c_{v}=c_1a_o/a_c,
\label{eq:cvir}
\end{equation}
where $c_1=4.1$, and $a_o$, $a_c$ are the scale factors at the time of 
observation and formation, respectively, a relation which (according to 
Wechsler et al.) fits well in the entire mass range explored in their 
simulations. Note that these authors have defined the characteristic 
formation time as the epoch in which the mass accretion rate falls below 
some specified value. This differs from the formation time definition in 
the LC formalism, where it is defined as the time at which a constant 
fraction $f$ of the halo mass has been accumulated. To assess the impact 
of the different definition of the formation time, Wechsler et al. 
repeated the analysis for merger trees produced by the ESF and compared 
the resulting halo distribution as function of the formation epoch $a_c$. 
Their results suggest that the ESF formalism predicts formation times 
higher (i.e. later) by a factor 1.25. Therefore, in order to use the 
scaling described in equation~(\ref{eq:cvir}) we should first divide the 
the ESF-based formation times indicated by this factor:
\begin{equation}
a_c=\frac{a_c^{EPS}}{1.25}=\frac{a_{obs}}{1.25(1+z_f)},
\end{equation}
where $z_f$ is the formation redshift. Using relation~(\ref{eq:cvir}) we 
finally obtain
\begin{equation}
c_{v}=5.125\frac{1+z_f}{1+z_{obs}},
\label{eq:cv2}
\end{equation}
where $a_{obs}\equiv a_0/(1+z_{obs})$. A PDF of the halo concentration 
parameter can now be derived by using the usual probability distribution 
transformation law, $P(z_f)\,dz_f=P(c_v)\,dc_v$, such that
\begin{equation}
P(c_v)=P(z_f)\left\vert\frac{dz_f}{dc_v}\right\vert=
\frac{1+z_{obs}}{5.125}P(z_f),
\end{equation}
whereas the corresponding CPDF - i.e. the probability that a halo has a 
concentration parameter higher than a given value - is calculated as 
$\int_{c_v}^{\infty}P(c_v')\,dc_v'$.

\section{Results}
\label{sec:results}

\begin{figure}
\centering
\epsfig{file=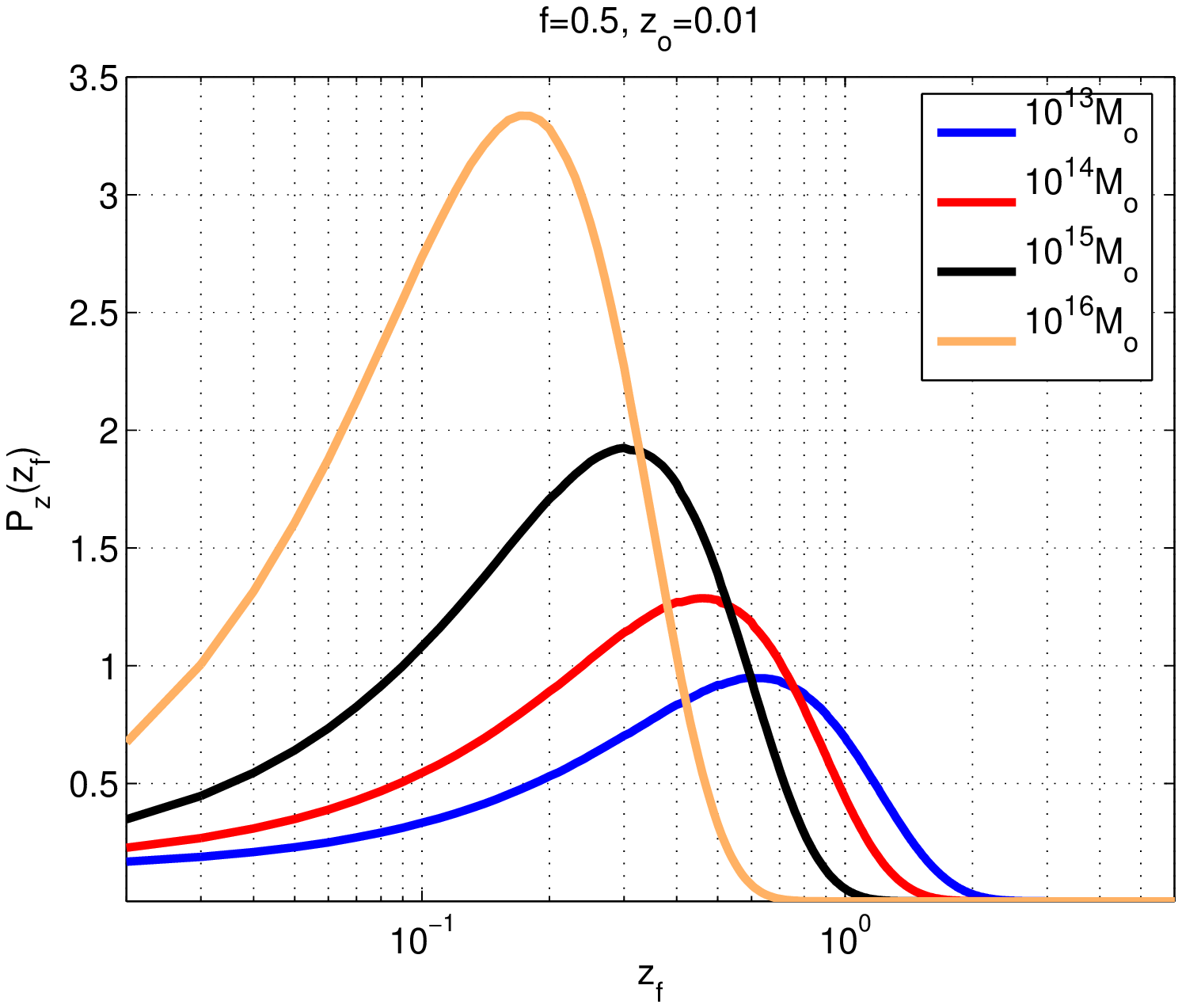, height=6.5cm, width=8.5cm, clip=}
\epsfig{file=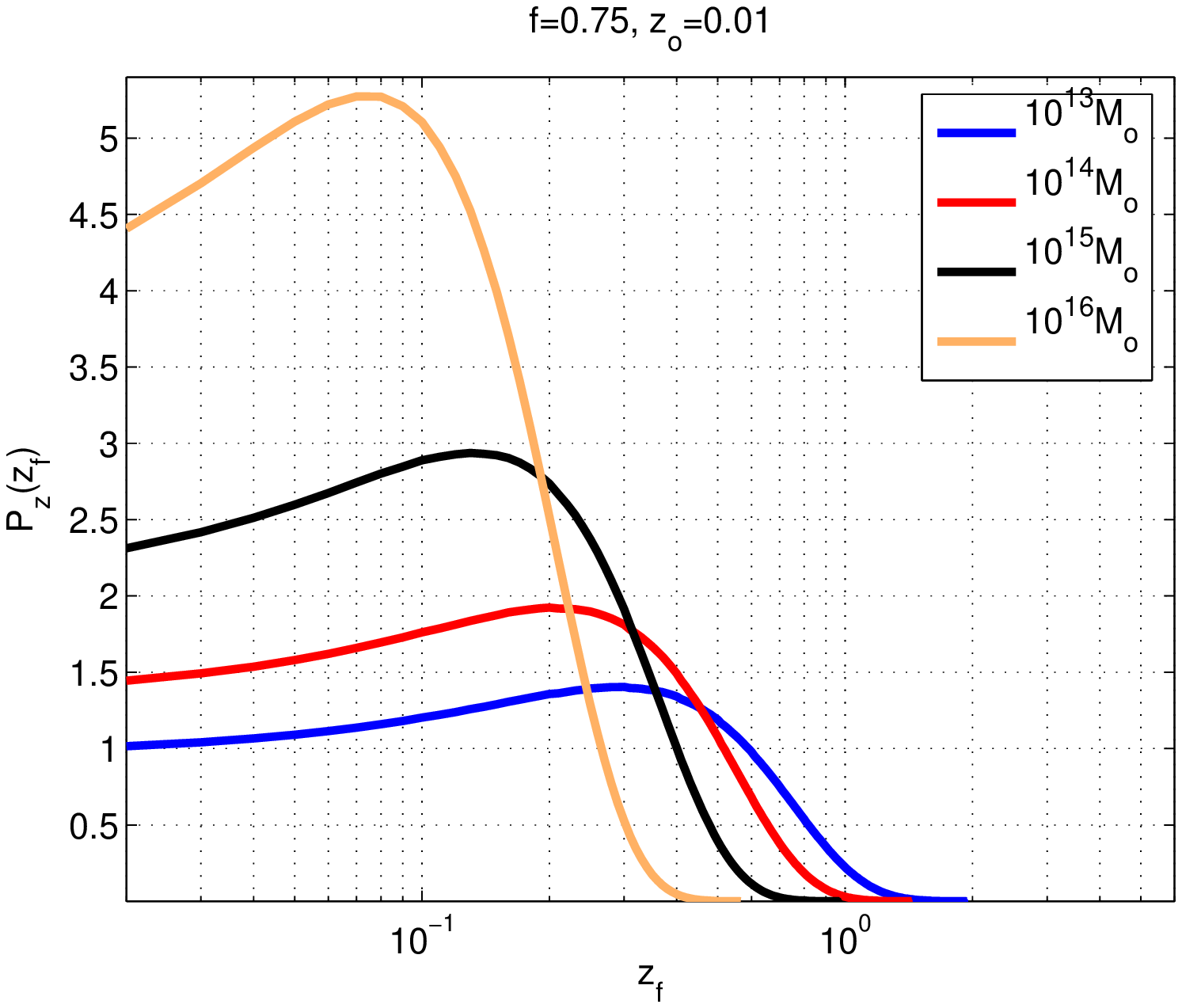, height=6.5cm, width=8.5cm, clip=}
\caption{PDF of formation times for the standard $\Lambda$CDM model. 
The blue, red, black and orange curves correspond to halo masses of 
$10^{13}\,M_{\odot}$, $10^{14}\,M_{\odot}$, $10^{15}\,M_{\odot}$, and 
$10^{16}\,M_{\odot}$, respectively, observed at redshift $z_{obs}=0.01$. 
The left- and right-hand panels correspond to a mass fraction of $f=0.5$ 
and $f=0.75$, respectively.}
\label{fig:ftdist}
\end{figure}

PDF of formation times of haloes with various masses at (observation) 
redshift $z_{obs}=0.01$ were calculated using equation~(\ref{eq:pz}), 
and are illustrated in Fig.~(\ref{fig:ftdist}). Haloes were defined to 
have formed once they assembled a fraction of either $f=0.5$ (left-hand 
panel), or $f=0.75$ (right-hand panel) of their current masses. In both 
panels of Fig.~(\ref{fig:ftdist}) blue, red, black and orange curves 
correspond to mass scales of $10^{13}, 10^{14}, 10^{15}$, and $10^{16}$ 
$M_{\odot}$, respectively. The merger picture describes the formation of 
haloes as an ongoing process of mutual collapse of subhaloes. Since low-mass 
subhaloes are more abundant than high-mass subhaloes, haloes of lower masses are more likely to form at higher redshifts. These gradually merge with other subhaloes to form increasingly larger structures, as can be clearly seen
in Fig.~(\ref{fig:ftdist}), where low-mass haloes ($10^{13}M_{\odot}$) 
form relatively early and at high abundances. The enhancement of the 
formation time probability and the consequent higher abundances. 
of subhaloes of this mass scale, imply higher likelihood for a merger to 
occur and higher-mass objects to form. The latter become progressively 
more common, and their active formation time - defined as the epoch at 
which the formation time PDF attains a maximum - peaks at a lower 
redshift with respect to the active formation epoch of their progenitors. 

High abundance of the lowest-mass subhaloes enhances the merger 
probabilities; as a result these subhaloes grow less common with time. 
Consequently, the PDF of their formation time is reduced with 
decreasing redshift, and is eventually dominated 
by the PDF of their higher-mass products. The process goes on, with 
ever-increasing mass objects becoming more abundant, accompanied by a 
reduction of the PDF for lower-mass haloes. This is apparent in either 
left-hand panel of the figure, where the PDF of formation times peak at 
lower redshifts with increasing mass. The fall of the curves from the 
active formation time towards lower redshifts can thus be explained 
by the fact that low-mass subhaloes hierarchically merge to yield 
higher-mass subhaloes, thereby diminishing the probability of 
low-mass objects to form at low redshifts. The corresponding 
progression in the high-mass halo range can be attributed to the fact 
that merger rates are substantially reduced with increasing mass. This, 
too, is in accord with expectations from the ESF, as was demonstrated by, 
e.g., LC.

The earlier formation times implied by the $f=0.5$ case with respect to 
the $f=0.75$ case are clearly reflected in the plots; the active 
formation time lies at manifestly higher redshifts in the $f=0.5$ case. 
Whereas it peaks at approximately $z=0.30, 0.20, 0.13, 0.07$ for haloes 
of $10^{13},10^{14}, 10^{15}, 10^{16}$ $M_{\odot}$ for $f=0.75$, the 
corresponding formation times with $f=0.5$ are $z=0.60, 0.46, 0.30, 
0.17$, i.e., at least twice as high. 
\begin{figure}
\centering
\epsfig{file=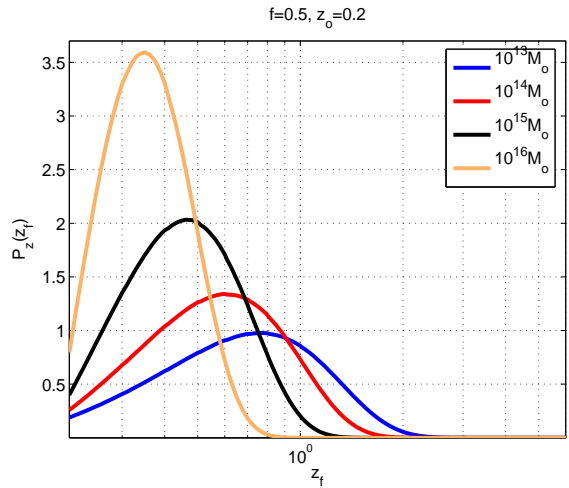, height=6.5cm, width=8.5cm, clip=}
\caption{The same as in Fig.~(\ref{fig:ftdist}), but for an observation
redshift of $z=0.2$. (Results shown are only for $f=0.5$.)}
\label{fig:ftdistrat}
\end{figure}
It can also be seen that the PDFs for the $f=0.5$ case have more 
pronounced peaks and fall more steeply on either side of the peak 
than for $f=0.75$, where the curves are decidedly 
flatter at lower redshifts than for $f=0.5$. This clearly is due to 
the fact that haloes can assemble half of their present mass at 
considerably earlier stages of their evolution than it takes to assemble 
3/4 of their present mass. Thus, their formation times peak at higher 
redshifts, and the likelihood of merging and forming subhaloes of higher 
masses increases at higher redshifts as well, resulting in lower 
probabilities for low-mass haloes to form at low redshifts. 
\begin{figure}
\centering
\epsfig{file=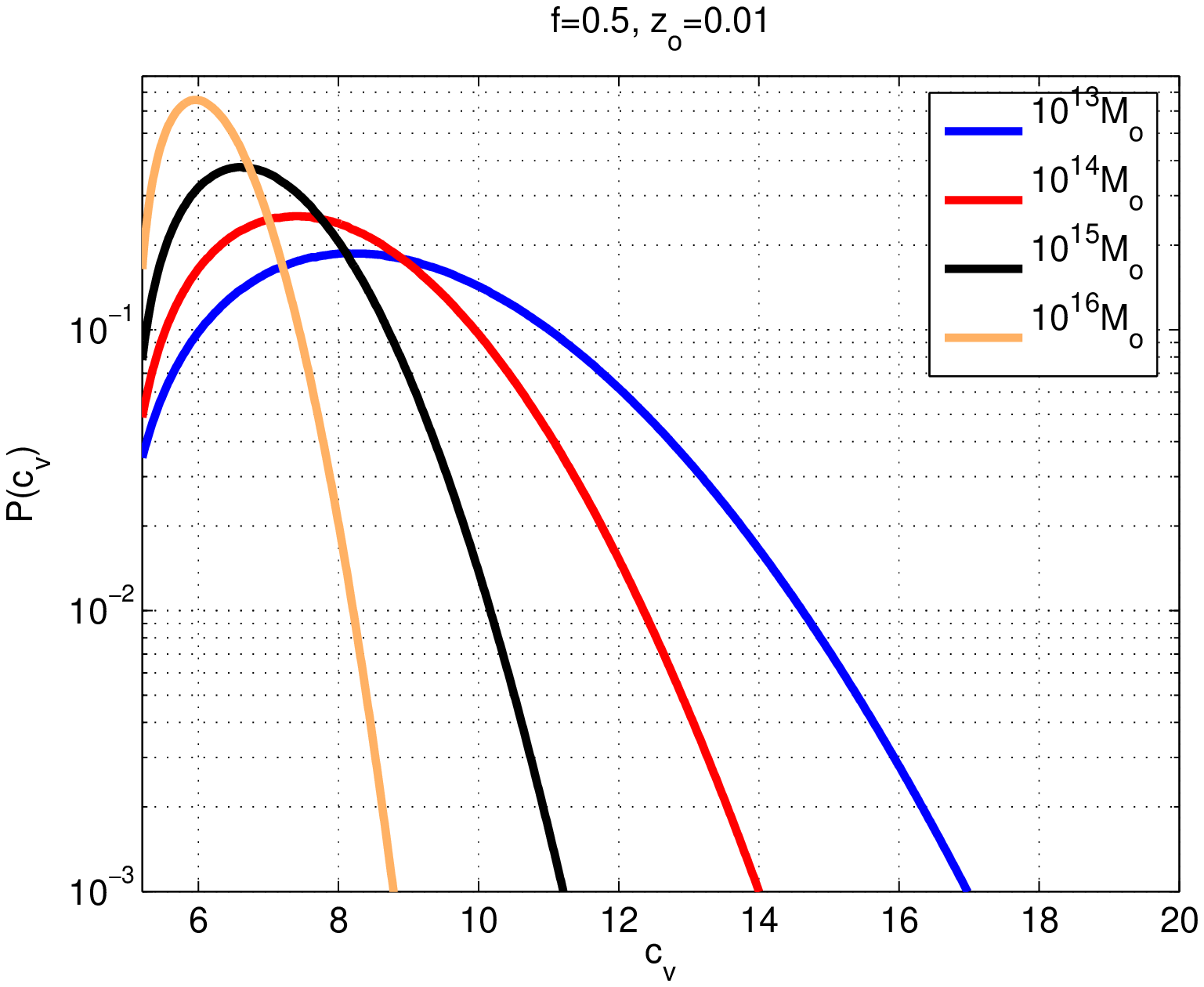, height=6.5cm, width=8.5cm, clip=}
\epsfig{file=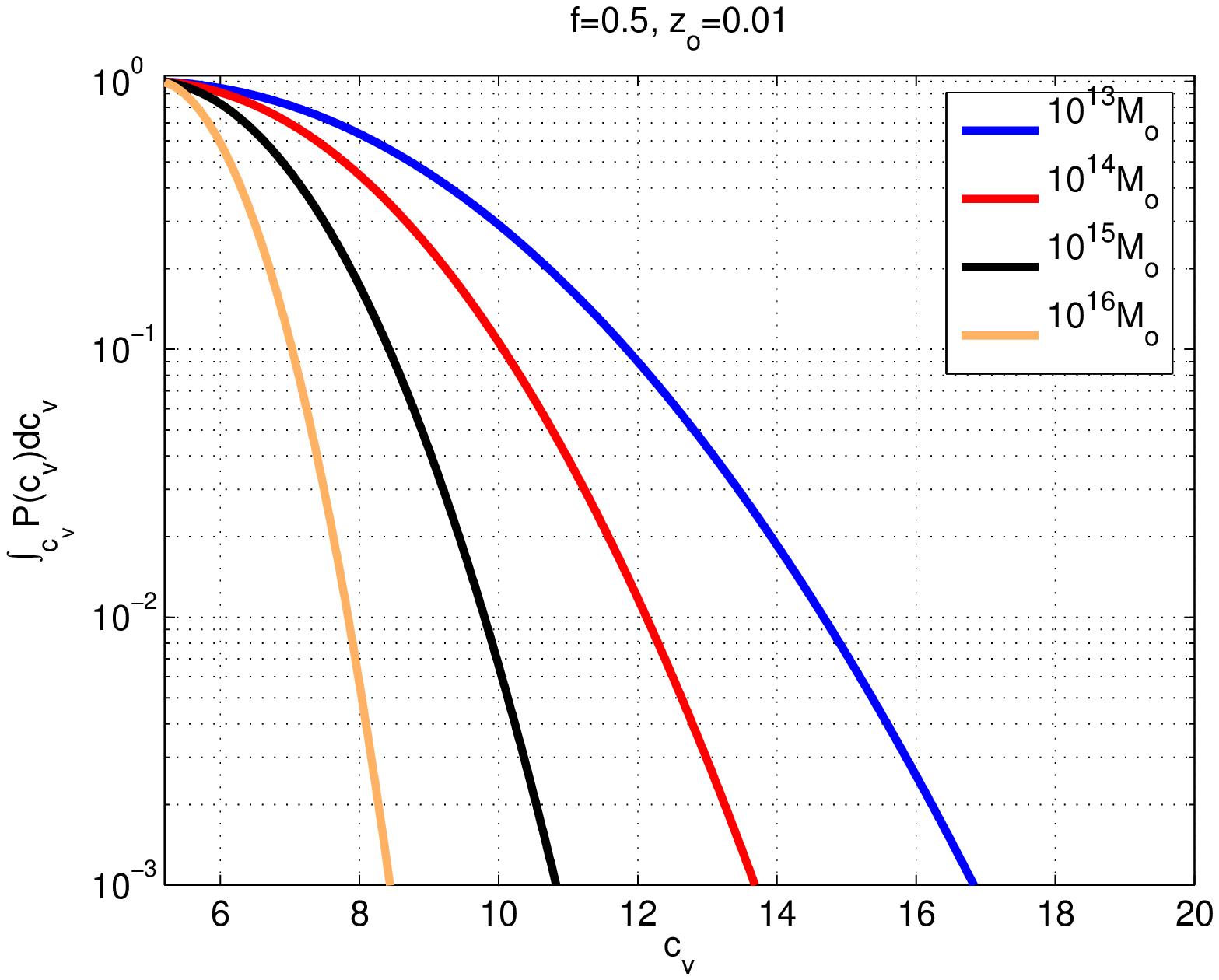, height=6.5cm, width=8.5cm, clip=}
\epsfig{file=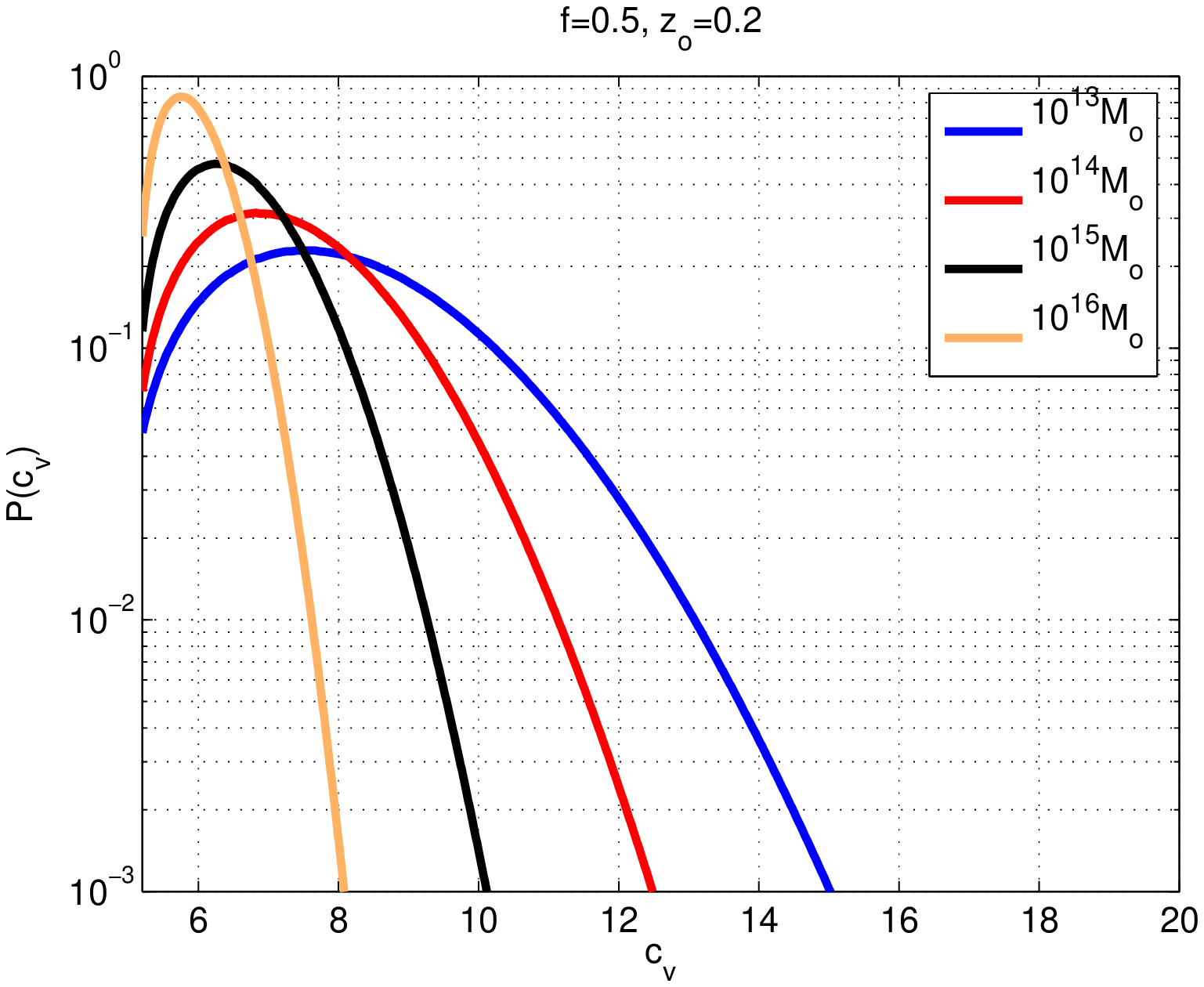, height=6.5cm, width=8.5cm, clip=}
\epsfig{file=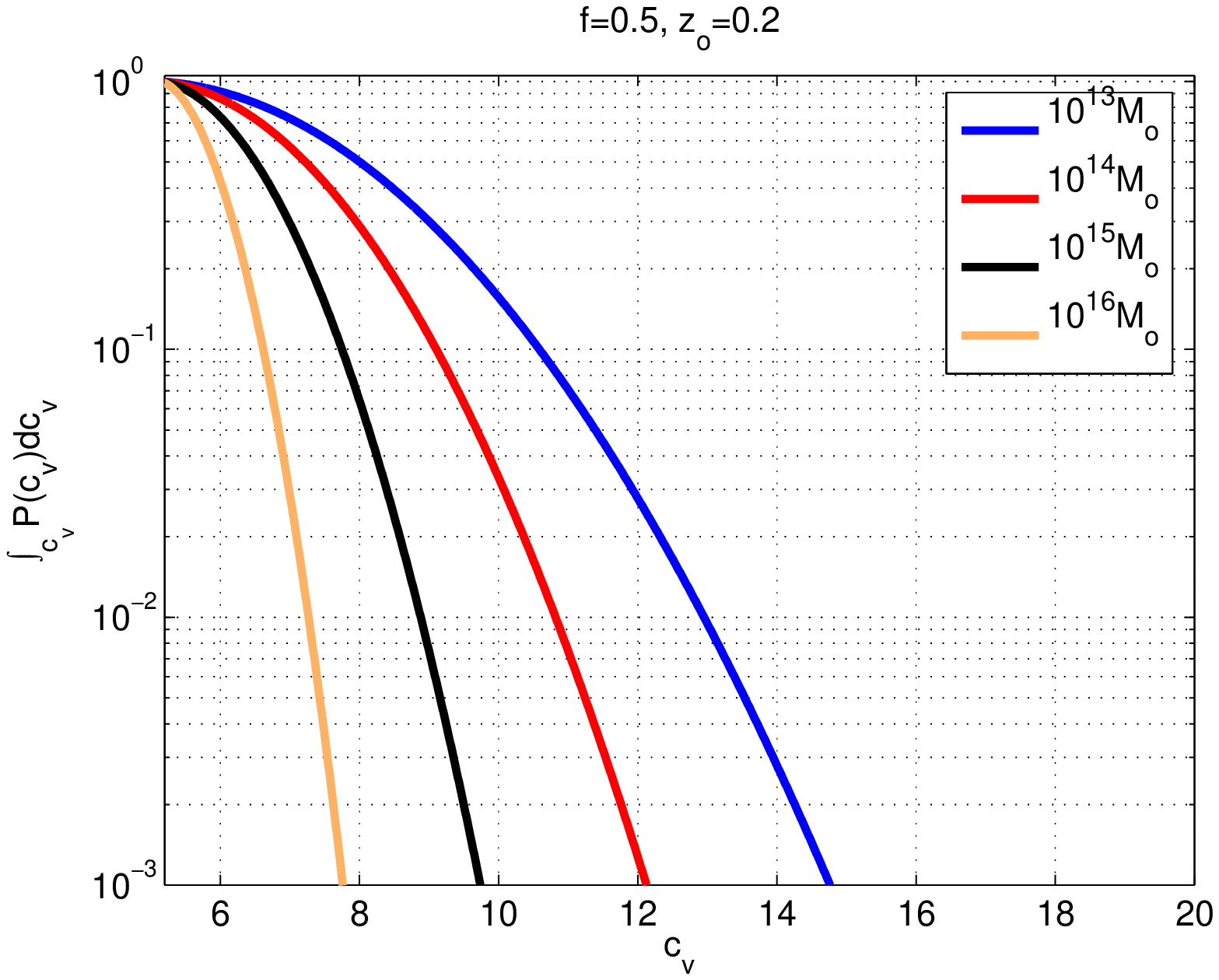, height=6.5cm, width=8.5cm, clip=}
\caption{The PDF (left) and CPDF (right) of the halo concentration 
parameter. Upper and lower panels correspond to observation redshifts 
of 0.01 and 0.2, respectively.}
\label{fig:cp}
\end{figure}
We have repeated the calculations for an observation redshift of 
$z_{obs}=0.2$ and $f=0.5$, results for which are illustrated in 
Fig.~(\ref{fig:ftdistrat}). Obviously, the PDFs shift to the right, 
towards higher redshifts, but retain their general shape.  

Results of the PDF and CDPF of halo concentrations are presented in 
Fig.~(\ref{fig:cp}) for observation redshifts of $z_{obs}=0.01$ and 
$z_{obs}=0.2$. For $z_{obs}=0.01$, haloes with 
$10^{13},\, 10^{14},\, 10^{15},\, 10^{16}$ $M_{\odot}$ 
peak at concentrations $c_v\sim 8.1,\, 7.4,\,  6.6,\, 6.0$, 
respectively. The corresponding peak concentrations for $z_{obs}=0.2$ 
are $c_v\sim 7.6,\, 7.2, \, 6.2, \, 5.8$, i.e. somewhat lower, since 
the higher observation redshift limits the redshift range available 
for halo formation. The CPDFs plotted in the right-hand panels of 
Fig.~(\ref{fig:cp}) provide a convenient means of determining the 
probability of finding a concentration higher than a given value. 
For example, at $z_{obs}=0.01$ the probability of finding haloes of 
$10^{13},\, 10^{14},\, 10^{15},\, 10^{16}$ $M_{\odot}$ 
with concentrations higher than $c_v=10$ are $\sim 0.29,\, 0.11,\, 0.007, 
\, 2.2\cdot 10^{-7}$. For $z_{obs}=0.2$ the corresponding values are 
$\sim 0.16,\, 0.03,\, 4.4\cdot 10^{-4},\, 4.0\cdot 10^{-11}$. Clearly, 
the probability for a high value of $c_v$ in a $\sim 10^{15} M_{\odot}$ 
cluster is quite low in the standard $\Lambda$CDM cosmology.

\section{The case of A1689}
\label{sec:er}

Recently, Broadhurst \& Barkanna (2007, hereafter BB) have reported the 
results of a comparative study of Einstein radii calculated for a large 
sample of simulated clusters of masses $\sim 10^{15}\,M_{\odot}$, and 
those observed in the three massive clusters A1689, C10024-17, and A1703. 
While the simulations yield Einstein radii in the range $\sim 15-25"$, the 
measured values for these clusters are $\sim 50", 31"$, and $32"$, 
respectively, a result which is seen by BB to constitute a challenge to 
the standard $\Lambda$CDM model. 
 
Since we know all relevant properties of A1689, namely its virial mass, 
virial radius, concentration parameter (assuming an NFW DM profile), 
redshift, and the lensed source redshift, it is possible to construct a 
PDF of Einstein radii for an A1689-like cluster. The equation governing 
the relation between the concentration parameter and the Einstein radius 
is, assuming an NFW profile (BB),
\begin{equation}
\left(\frac{4R_v\rho_c^z\Delta_c}{3\Sigma_{cr}}\right)
\frac{c_v^2}{\ln{(1+c_v)}-c_v/(1+c_v)}\frac{g(x)}{x^2}=1,
\label{eq:gx}
\end{equation}
where $R_v$, $\rho_c^z$, $\Delta_c$, and $\Sigma_{cr}$ are the virial 
radius, critical density at redshift $z$, overdensity at virialisation, 
and critical surface density, respectively. All these quantities are 
specified by BB, with the exception of the virial radius, for which we 
adopt the theoretical value found from the relation 
$M_v=\frac{4\pi}{3}\rho_c(z)\Delta_c(z)R_v^3$, resulting in 
$R_v=2.63\,Mpc$, for $h=0.687$. Also, $x\equiv\frac{R_E c_v}{R_v}$, 
where $R_E$ is the Einstein radius, and
\begin{eqnarray}
g(x)=\ln{\frac{x}{2}}+
\left\{\begin{array}{ll}
1, &\mbox{$x=1$} \\
\frac{2}{\sqrt{x^2-1}}\tan^{-1}\sqrt{\frac{x-1}{x+1}}, &\mbox{$x>1$} \\
\frac{2}{\sqrt{1-x^2}}\tanh^{-1}\sqrt{\frac{1-x}{1+x}}, &\mbox{$x<1$}
\end{array}\right..
\end{eqnarray}
The solution of Eq.~(\ref{eq:gx}) provides the Einstein radius as a function 
of $c_v$, from which it is trivial to derive the angular Einstein radius, 
$\theta_E=R_E/D_A(z_{l})$, the ratio between the (physical) Einstein 
radius and the angular diameter distance to the lens, the measured 
redshift of A1689, $z_l=0.183$. It remains to determine the PDF of the 
angular Einstein radius. This can be accomplished using once more the 
transformation law of probability distribution functions:
\begin{equation}
P(\theta_E)d\theta_E=P(c_v)dc_v=P(z_f)dz_f,
\end{equation}
from which we obtain
\begin{equation}
P(\theta_E)=P(c_v)\left\vert\frac{dc_v}{d\theta_E}\right\vert=
P(z_f)\left\vert\frac{dz_f}{dc_v}\right\vert
\left\vert\frac{dc_v}{d\theta_E}\right\vert.
\end{equation}
The first derivative is calculated analytically using Eq.~(\ref{eq:cv2}), 
whereas a numerical calculation based on the $\theta_E(c_v)$ relation in 
Eq.~(\ref{eq:gx}) is used in order to infer the second derivative. 

\begin{figure}
\centering
\epsfig{file=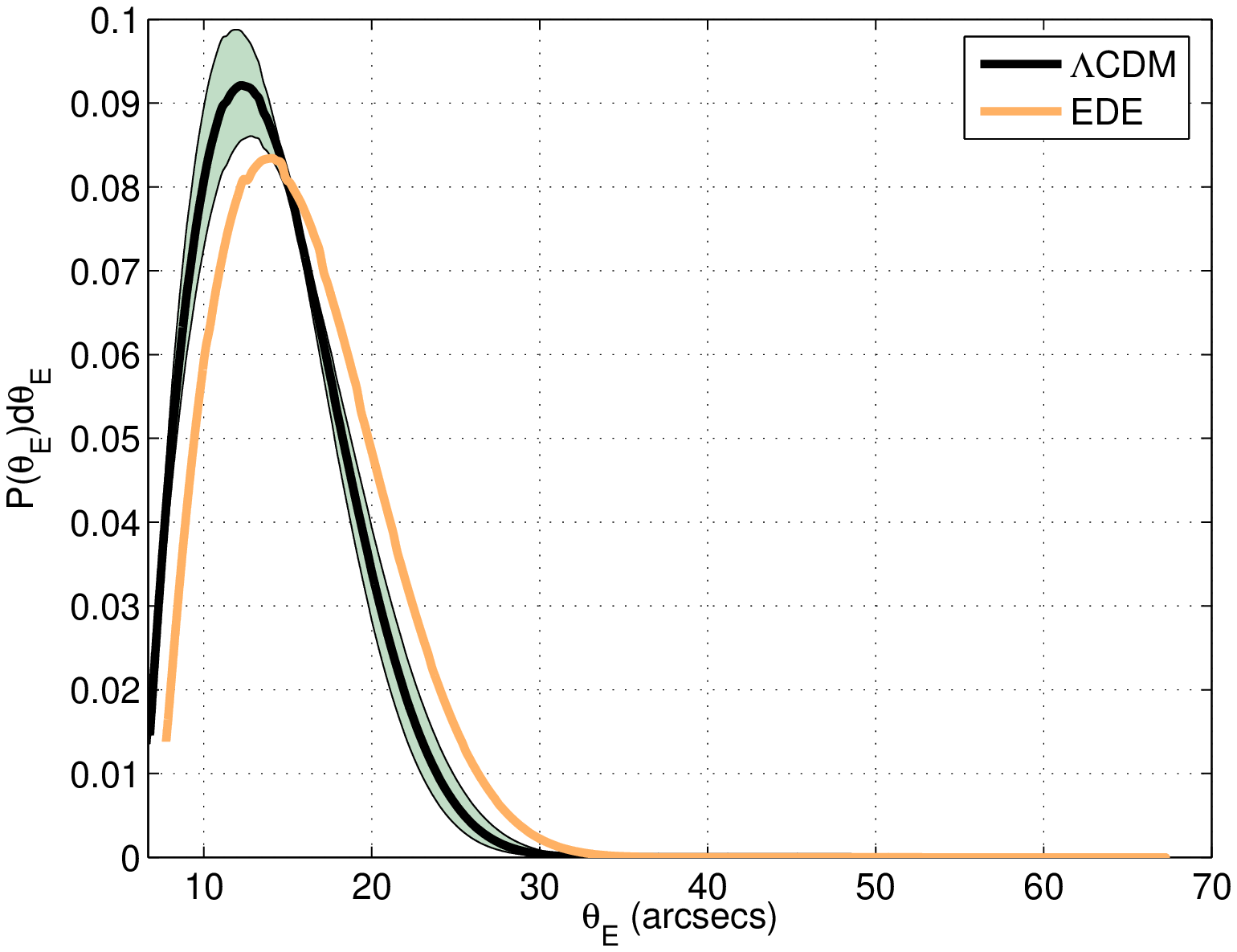, height=6.5cm, width=8.5cm, clip=}
\epsfig{file=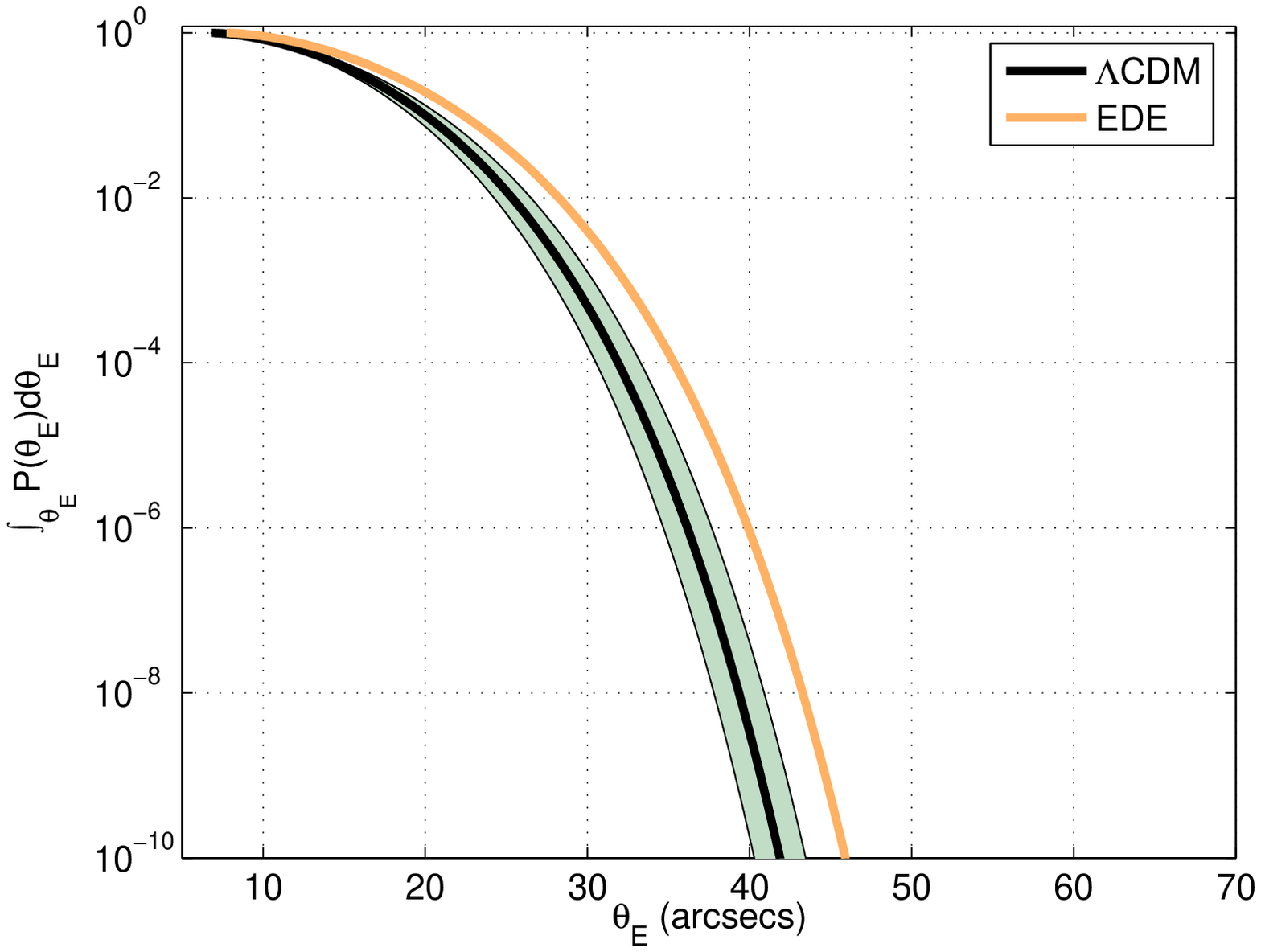, height=6.5cm, width=8.5cm, clip=}
\caption{The PDF (left) and CPDF (right) of the Einstein radius for an 
A1689-like cluster. Continuous black and orange curves correspond to the 
standard $\Lambda$CDM and EDE models, respectively. The shaded areas 
reflect the variance generated by incorporating the $2\sigma$ uncertainty 
in the cosmological parameters $h$, $n$, and $\sigma_8$ in the 
calculations for the $\Lambda$CDM model.}
\label{fig:er}
\end{figure}

Results for the PDF and CDF of the angular Einstein radius for an 
A1689-like cluster are presented in Fig.~(\ref{fig:er}). The results 
generated within the framework of the standard $\Lambda$CDM are 
represented by the black continuous curve; the shaded areas represent 
the $2\sigma$ uncertainties in the cosmological parameters 
$\Omega_m$, $n$, and $\sigma_8$. As is clear from the figure, large 
Einstein radii ($\gtrsim 40"$) have extremely low probabilities in the 
standard $\Lambda$CDM universe. In fact, the cumulative probability for 
an A1689-like cluster to induce an angular Einstein radius larger than 
$40"$ amounts to $\sim 3.4\cdot 10^{-9}$ in this model; the corresponding 
probability for the $+2\sigma$ level increases to $\sim 4.1\cdot 10^{-8}$.

Non-standard cosmological models characterised by earlier evolution of 
the large scale structure may resolve the apparent discrepancy between 
measured and predicted values of $R_E$, by virtue of earlier halo 
formation times, reflected in higher central halo concentrations and, 
therefore, higher probabilities for large Einstein radii. For example, 
models based on positively skewed primordial density fluctuations lead 
to earlier growth of the large scale structure. Early dark energy 
(hereafter EDE) models provide another alternative for inducing earlier 
formation times by virtue of the modified cosmic dynamics implied by a 
non-negligible DE component in the early universe. To explore this 
possibility, we have repeated our calculations for a specific EDE model, 
with an early quintessence density parameter $\Omega^{e}_d=0.03$ and 
equation of state coefficient $w_0=-0.9$ at $z=0$. These parameters are 
consistent with recent WMAP results (Doran \& Robbers, 2006). The other 
cosmological parameters were not modified, with the exception of 
$\sigma_8=0.51$, a value obtained by normalizing the cumulative halo 
mass function to the same number of halos generated in the standard 
$\Lambda$CDM model. Complete details concerning the calculation of the 
relevant large scale quantities within the framework of EDE models can be 
found in, e.g., Bartelmann, Doran, \& Wetterich (2006) and Sadeh, Rephaeli 
\& Silk (2007). 
The latter work was motivated by the possible need to have a higher level 
of CMB anisotropy induced by the Sunyaev-Zeldovich effect than is predicted 
in the standard $\Lambda$CDM model.

Results of the PDF and CDF of the Einstein radius for A1698 are shown by 
the continuous orange curve of Fig.~(\ref{fig:er}). Note that for the 
EDE model the theoretically inferred virial radius reduces slightly to 
$r_v=2.54\,Mpc$. The probability for an A1689-like cluster to produce an 
Einstein radius larger than $40"$ in this model is $\sim 9.1\cdot 
10^{-7}$, i.e., approximately two orders of magnitude higher than the 
corresponding probability in the $\Lambda$CDM model. The $+2\sigma$ 
uncertainty (which is not shown in the plot) in the cosmological 
parameters would lead to even higher probabilities than those indicated 
by the orange curve of the EDE model. However, these probabilities are 
still very low.

\section{Discussion}
\label{sec:disc}

The observationally inferred high concentration parameters associated 
with several massive clusters have been claimed to constitute 
a major challenge to the currently favoured standard cosmological model. 
Such high concentrations are expected in haloes formed at relatively high 
redshifts, reflecting the high universal background density at that time. 
As Einstein radii in lensing haloes increase monotonically with their 
concentration parameters, observed high radii also indicate 
earlier formation times. Massive clusters form preferentially at low 
redshifts, as attested by results of both N-body simulations and 
semi-analytic calculations based on the excursion set theory. 

Our assessment of the apparent discrepancy between theoretical predictions 
and measurements of the concentration parameter and Einstein radii of 
high-mass clusters has been based on the assumption of NFW mass profiles. 
This is motivated by the explicit use of this profile in the analysis 
of N-body simulations (Wechsler et al. 2002, Neto et al. 2007), which were 
used to derive the scaling relation between $c_v$ and $t_f$ 
equation~(\ref{eq:cv2}) in these works, and had to be adopted in our 
analysis for consistency. The NFW profile is not the only profile that 
provides an acceptable fit to simulated haloes. It is quite possible that 
most of the discrepancy explored here stems from the insistence on 
using the NFW profile. Even so, the fact that this profile is used 
extensively in the description of galaxies and clusters provides 
sufficient motivation to focus on its implications in the context of 
our work here.

If indeed the apparent discrepancy is considered a serious difficulty 
for the standard model, then it is only reasonable to consider close 
alternatives to the standard $\Lambda$CDM model. Earlier 
formation times are naturally predicted in non-standard cosmologies, 
such as those characterised by non-Gaussian, positively skewed primordial 
density fluctuations, and early dark energy models. We have shown in 
this work that cosmological models incorporating an early DE component 
may induce higher probabilities of finding large Einstein radii in 
A1689 than in the standard $\Lambda$CDM model, although perhaps not 
sufficiently high to remove the apparent conflict between theory and 
observations.

\begin{figure}
\centering
\epsfig{file=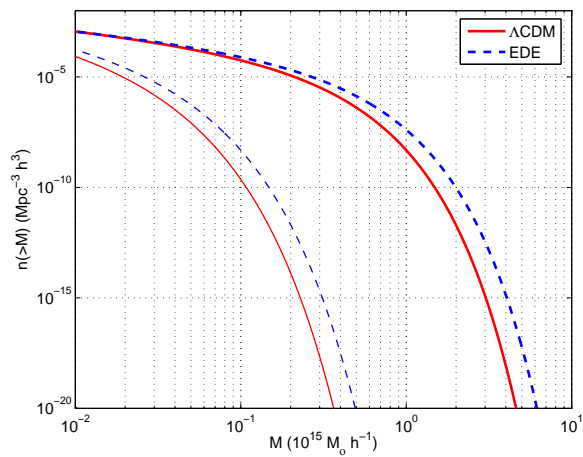, height=6.5cm, width=8.5cm, clip=}
\caption{The cumulative PS mass functions for the standard $\Lambda$CDM (red continuous curve) and EDE (blue dashed curve) models. Thick and thin lines correspond to observation redshift $z_{obs}=0.01$ and $z_{obs}=3$, 
respectively.}
\label{fig:mf}
\end{figure}

The calculation of the number of clusters whose masses and redshifts 
are comparable to or larger than those of A1689, and which are expected 
to induce Einstein radii larger than a given value, is not simple since 
this would require an integration over the formation time PDFs of all 
of these masses and redshifts. In addition to this technical issue, one 
would also need to draw an arbitrary source redshift so as to be able to 
calculate the corresponding angular diameter distances needed in 
Eq.~(\ref{eq:gx}). However, we can provide a crude estimate of the upper 
limit to this number by integrating the relevant PS mass functions - 
shown in Fig.~(\ref{fig:mf}) for both models at observation redshifts 
$z_{obs}=0.01$ and $z_{obs}=3$ - over the mass range [$10^{15}\,M_{\odot},
\infty$].

With our specific choice of cosmological parameters, the standard 
$\Lambda$CDM and EDE models described in this work predict $\sim 50$ 
($\sim 157$ for the $+2\sigma$ level of $\sigma_8$ - which agrees 
better with the corresponding figure inferred from large scale 
studies), and $\sim 300$ cumulative cluster counts with 
$M\ge 10^{15}\,M_{\odot}$ and $z_{obs}\ge 0.183$, respectively. 
Obviously, not all of these clusters have Einstein radii larger than $40"$, 
as these depend also on the redshifts of the lensed objects. We can readily 
set an upper limit on the number of clusters with $\theta_E\ge 40"$, whose 
mass and redshift are within these ranges, by multiplying the cumulative 
numbers by the probabilities of finding an A1689-like cluster with 
$\theta_E\ge 40"$. As specified in the previous section, these probabilities 
are $3.4\cdot 10^{-9}$ (or $4.1\cdot 10^{-8}$ at the $+2\sigma$ 
level), and $9.1\cdot 10^{-7}$ in the standard $\Lambda$CDM and EDE models, 
respectively. Thus, the estimated numbers of clusters are 
$\sim 1.7\cdot 10^{-7}$ ($6.4\cdot 10^{-6}$ at the $+2\sigma$ level) and 
$\sim 2.7\cdot 10^{-4}$ in the $\Lambda$CDM and EDE models, respectively. 

It is important to realize that while our analysis of the PDF of the 
concentration parameter and Einstein radius poses an appreciable 
difficulty for the standard large scale model, it cannot be construed 
as insurmountable evidence against the validity of the standard 
$\Lambda$CDM model. In fact, there is a significant level of uncertainty 
in the relation between the formation time and concentration parameter, an 
uncertainty which we have not yet taken explicitly into account. 
Many factors contribute to this uncertainty, including the reported 
variance in the scaling, as shown by Wechsler et al. to be at the level 
of $\sim 50\%$, such that, for example, a halo with mass 
$10^{14}M_{\odot}h^{-1}$ would have a concentration $c_v\sim 7\pm 3$, 
representing only the $1\sigma$ uncertainty level. The actual uncertainty 
may even be appreciably higher, due to the fact that the baryonic component 
was ignored in the cited N-body simulations, and due to the large degree or 
arbitrariness in the definition of the formation time. 

The higher than predicted value of $\theta_E$ is reflected more directly in 
the discrepancy between the theoretically predicted and measured values of 
$R_v$. The solutions of Eq.~(\ref{eq:gx}), which yield the Einstein radius as a function of concentration parameter, are very sensitive to the virial 
radius of the cluster. We have used the most probable theoretically-inferred 
values of $R_v$ in our calculations, $R_v=2.63$ and $2.55\,Mpc$ in the 
standard $\Lambda$CDM and EDE models, respectively. But these values too 
could be uncertain; we explore here the impact of a variance of $\sim 20\%$ 
in $R_v$ that would presumably reflect the intricacies of the process of 
hierarchical clustering through mergers and non-spherical collapse, which 
was shown (e.g. Del Popolo 2002) to yield significantly lower virialisation 
overdensities than predicted in the spherical collapse scenario. For example, 
an isolated prolate spheroid of axial ratio $a_3/a_1=2.4:1$, which is the 
mean ratio found in an N-body simulation in which 878 clusters were 
identified (Hennawi et al. 2007), was shown by Del Popolo to have a 
virialisation overdensity of $\Delta_c\sim 50$, lower by more than a factor 
$3$ than the corresponding quantity for a spehrical virialised region. 
(Although carried out in the context of an Einstein de Sitter universe, the 
reported results are not likely to change qualitatively in a 
$\Lambda$CDM model.) For a given mass and redshift, 
$M_v=\frac{4\pi}{3}\rho_c(z)\Delta_c(z)R_v^3$, the lower overdensities 
obviously imply larger virial radii. 

\begin{figure}
\centering
\epsfig{file=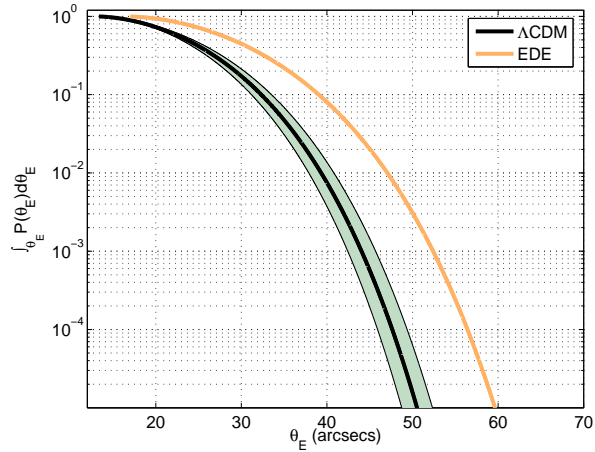, height=6.5cm, width=8.5cm, clip=}
\caption{The same as Fig.~(\ref{fig:er}), but for $R_v=3.1\,Mpc$.}
\label{fig:er2}
\end{figure}

A larger virial radius is also in accord with the results of a recent detailed, model-independent analysis of A1689 by Lemze et al. (2008), who (deduced the mass profile and) determined $R_v=2.14\,Mpc\cdot h^{-1}$ from joint fitting to extensive X-ray and lensing measurements. With $h=0.687$ we have $R_v=3.1\,Mpc$, which is 
indeed higher by $\sim 20\%$ than the theoretically predicted value for 
this cluster. This has a very significant impact on the PDF and CPDF of 
the Einstein radius, as demonstrated in Fig.~(\ref{fig:er2}). 

With this larger virial radius, the probability for an A1689-like cluster 
to have $\theta_E \ge 40"$ is $8\cdot 10^{-3}$ (or $1.3\cdot 10^{-2}$ at the $+2\sigma$ level) in the standard $\Lambda$CDM model, and $8\cdot 10^{-2}$ in the EDE model. These large differences originate in the larger Einstein radii predicted by Eq.~(\ref{eq:gx}) for the same concentration parameter, and give rise to substantially higher upper limits on the number of clusters expected to be detected with $\theta_E\ge 40"$, $\sim 1.3$ (at the $+2\sigma$ level), and as many as $\sim 24$ in the standard $\Lambda$CDM and EDE models, respectively. 

Finally, we note that the above considerations indicating significantly 
higher halo concentrations and larger Einstein radii in clusters may have 
significant ramifications for the non-standard, EDE model: Had the 
simulations been carried out within the framework of a specific EDE 
model, the formation time - concentration scaling would have probably 
predicted higher concentrations, by virtue of the earlier formation of 
structure. Here, too, the probabilities of finding haloes with high 
concentrations and Einstein radii would be markedly higher, perhaps even 
prohibitively higher, so much so that this could turn out to be a significant constraint on the parameters of this model.

\section{ACKNOWLEDGMENT}

Useful discussions with Dr. Tom Broadhurst are gratefully acknowledged. This work was supported by a grant from the Israel Science Foundation.

\section{References}
\def\ref{\par\noindent\hangindent 20pt}

\ref Bartelmann M., Doran M., Wetterich C., 2006, \aa, 454, 27
\ref Bond J.R., Cole S., Efstathiou G., \& Kaiser N., 1991,
\apj, 379, 440
\ref Broadhurst T.J., Barkanna R., 2008, preprint (astro-ph/0801.1875)
\ref Bullock J.S., Kolatt T.S., Sigad Y., Somerville R.S., Kravtsov A.V., Klypin A.A., Primack J.R., Dekel A., 2001, \mn, 321, 559
\ref Del Popoplo A., 2002, \aa, 387, 759
\ref Doran M., Robbers G., 2006, \jca, 6, 26
\ref Jing Y.P., 2000, 535, 30
\ref Lacey C., Cole S., 1993, \mn, 262, 627
\ref Lemze D., Barkana R., Broadhurst T.J., Rephaeli Y., 2007, preprint (astro-ph/0711.3908)
\ref Navarro J.F., Frenk C.A., \& White S.D.M, 1995, \mn, 275, 720
\ref Neto A.F., Gao L., Bett P., Cole S., Navarro J., Frenk C.S., White S.D.M, Springel V., Jenkins A., 2007, \mn, 381, 1450
\ref Press W.H. \& Schechter P., 1974, \apj, 187, 425
\ref Sadeh S., Rephaei Y., Silk J., 2007, \mn, 380, 637
\ref Spergel D.N. et al. 2007, \apjs, 170, 377
\ref Wechsler R.H., Bullock J.S., Primack J.R., Kravtsov A.V., \& Dekel A.,
2002, \apj, 568, 52
\ref Zhao D.H., Mo H.J., Ping Y.P., \& B\"{o}rner G., 2003, \mn, 339, 12

\bsp
\label{lastpage}

\end{document}